\documentclass[11pt,a4paper]{article}
\usepackage{graphicx}
\usepackage{amsmath} %for arrows
\usepackage{amssymb} %for arrows
\usepackage{mathtools} %for harpoons
\usepackage{xcolor}
\usepackage{overpic}
\usepackage{tabularx}

\usepackage[numbers]{natbib}

\usepackage[a4paper]{geometry}% or [a4paper,left=3cm,right=2cm,top=2.5cm,bottom=2.5cm]

\newcommand{\pers}[0]{s^{-1}}
\newcommand{\perM}[0]{\mu\text{M}^{-1}\pers}
\newcommand{\C}[1]{{#1}}
\newcommand{\REV}[1]{{#1}}
\newcommand{\REVV}[1]{{#1}}

\begin{document}

\title{A Mathematical Framework for Kinetochore-Driven Activation Feedback in the Mitotic Checkpoint}
\date{}
\maketitle

\author{{\center \bf Bashar Ibrahim } \\
              Department of Mathematics and Computer Science, University of Jena, Ernst-Abbe-Platz 2, 07743 Jena \\
}

\maketitle

\begin{abstract}

\C{Proliferating cells properly divide into their daughter cells through a process
that is mediated by kinetochores, protein-complexes that assemble at the
centromere of each sister chromatid.}
Each kinetochore has to establish a tight bipolar attachment
to the spindle apparatus before sister-chromatid separation is initiated.
\C{The} Spindle Assembly Checkpoint (SAC) links the biophysical attachment
status of the kinetochores to mitotic progression, \C{and ensures that even a single
misaligned kinetochore keeps the checkpoint active.}
\C{The }mechanism by which this is achieved is still elusive.
Current computational models of the human SAC disregard important biochemical properties
\C{by omitting \REV{any kind of feedback loop, proper kinetochore signals,
and other spatial properties such as the stability of the system and diffusion effects.}}
To allow for \C{more realistic} \emph{in silico} study of the dynamics of \C{the SAC model},
\REV{a minimal mathematical framework for SAC activation and silencing
is introduced.
\C{A} nonlinear ordinary differential equation model successfully
\C{reproduces bifurcation signaling switches with attachment of all 92
kinetochores and activation of APC/C by kinetochore-driven feedback.}
}
\C{A} partial differential equation model and mathematical linear stability \C{analyses}
\C{indicate the influence of diffusion and system stability.}
The conclusion is that quantitative models of the human SAC \C{should} account for
the \REV{positive feedback on APC/C activation driven by the kinetochores which is essential for SAC silencing.
Experimental diffusion coefficients for MCC sub-complexes are \C{found to be} insufficient for
rapid APC/C inhibition.
The presented analysis allows for systems-level understanding \C{of} mitotic control
and the minimal new model can function as a \C{basis} for developing further quantitative-integrative models
\C{of} the cell division cycle.}
\end{abstract}

\newpage
\section{Introduction}
\label{intro}
Monitoring the fidelity of chromosome segregation during \C{the} cell life cycle
relies on transition control mechanisms, called checkpoints,
\C{which ensure} that all criteria are met before moving on irreversibly to the next phase \cite{Musacchio2007,Ibrahim2015Review}.
The major control mechanism in mitosis is called the Spindle Assembly Checkpoint (SAC; sometimes referred to as the mitotic checkpoint \cite{Rudner1996}).
The SAC ensures that all chromosomes are properly attached to spindle microtubules through their kinetochores.
Even \C{a single} unattached or mis-attached kinetochore (out of 92 in \C{a} human cell) is sufficient to keep the SAC engaged, yet the
mechanism by which this is achieved is still elusive \cite{Rieder1994,Rieder1995}.
A malfunction in \C{the} SAC process can cause aneuploidy and lead to tumorigenesis \cite{Morais2013,Holland2009,denisenko2016}.

\begin{figure}[htp]
\begin{center}
\includegraphics[width=\textwidth]{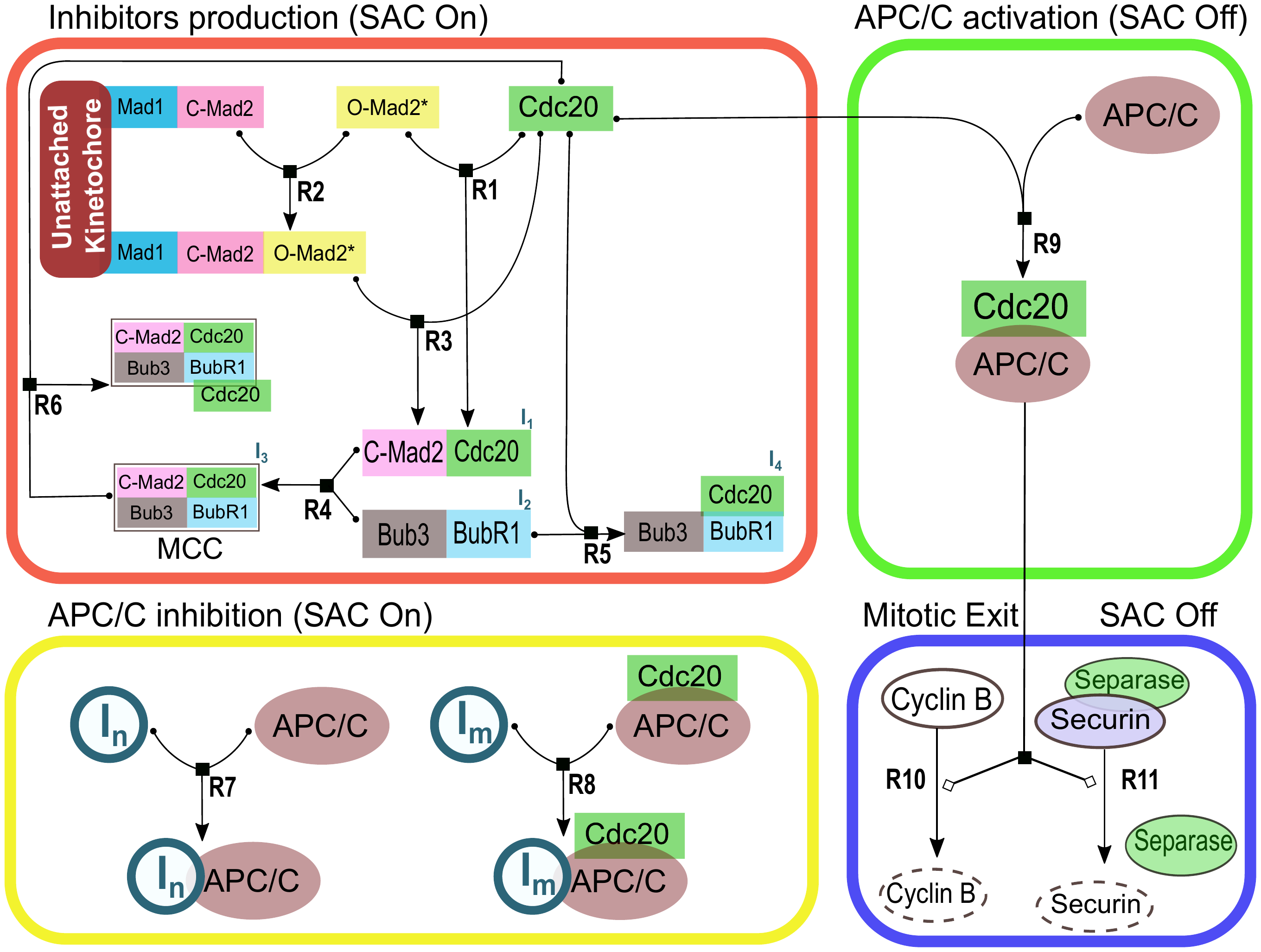}
\caption{Schematics illustrating the intracellular signaling of spindle assembly checkpoint activation and silencing.
    %
    % JAL - Perhaps 'Inhibitors producing' -> 'Inhibition' or 'Inhibitor production'?
    \emph{\REV{Inhibitors production} (red box):}
    The protein Mad2 is present in two stable conformations
    differing in the spatial arrangement of a `safety-belt' that is
    either open (O-Mad2) or closed (C-Mad2) \cite{DeAntoni2005,Luo2004}.
    O-Mad2 is able \C{to transiently}
    bind Cdc20 (R1).
    Meanwhile, O-Mad2 is recruited to unattached kinetochores
    by Mad1-C-Mad2 to form the ternary complex Mad1:C-Mad2:O-Mad2* \cite{DeAntoni2005,Luo2004},
    which can bind Cdc20 efficiently, and switches to \C{a}
    closed conformation upon Cdc20-binding (R2-R3).
    The resulting complex, Cdc20:C-Mad2, together with BubR1:Bub3, forms the tetrameric
    mitotic checkpoint complex (MCC, R4), which is
    a potent inhibitor of APC/C. The
    Cdc20 \C{also binds the} BubR1:Bub3 complex independently (R5).
    MCC, which contains Cdc20 as a subunit, can bind an additional free Cdc20 (R6).
    \emph{APC/C inhibition (yellow box):} The symbol $\textbf{I}_\textbf{n}$ refers to APC/C potential inhibitors (cf. MCC, Mad2:Cdc20, bubR1:bub3, and MCF2; see R7).
    The $\textbf{I}_\textbf{m}$ symbol denotes the ability of \C{the MCC complex to additionally inhibit Cdc20 that is already} bound to APC/C (R8).
    \emph{APC/C activation (green box):}
    When all 92 kinetochores are correctly attached from opposite poles to the mitotic spindle,
    \C{the SAC is switch off and APC/C is activated by Cdc20 via the} APC/C:Cdc20 complex (R9).
    \emph{Mitotic Exit (blue box):}
    Active APC/C \C{contributes to the} degradation of mitotic
    Cyclin B (R10), and causes Securin (budding yeast Pds1) to be
    tagged for degradation by the \C{proteasome, making} Seprase fully active (R11).
    }
    \label{fig:sacnetwork}
\end{center}
\end{figure}

Biochemically, it is thought that unattached or misaligned kinetochores catalyze the formation
and broadcasting \C{of} a `wait' signal to the environment (cf. Fig. \ref{fig:sacnetwork}, R1-R5). \C{This counters} the
activation of the ubiquitin ligase anaphase promoting complex/cyclosome (APC/C) \C{by Cdc20}.
APC/C activity is \C{thought} to be inhibited in multiple ways.
\C{It can be} directly prevented by \C{a potent inhibitor}, the Mitotic
Checkpoint Complex (MCC), which consists of the four checkpoint
proteins Mad2, BubR1, Bub3, and Cdc20 (cf. Fig. \ref{fig:sacnetwork}, R7-R8).
Furthermore, \C{it has been suggested that} MCC sub-complexes (\C{such as} BubR1 and Mad2:Cdc20) \C{interact}
with APC/C \cite{Fang1998,Han2013}. Another inhibitor,
called the mitotic checkpoint factor 2 (MCF2), is associated
with APC/C in the checkpoint arrested state but its
composition is \C{unknown} \cite{Eytan2008} (cf. Fig. \ref{fig:sacnetwork}, R7).
Recent compelling data shows that the MCC itself can bind to Cdc20 that has already bound to APC/C or \C{to}
a free Cdc20 \cite{Izawa2015} (cf. Fig. \ref{fig:sacnetwork}, R6).
With the exception of MCF2, all complexes inhibiting APC/C rely
on the presence \C{of} unattached kinetochores for \C{sufficiently rapid} formation \cite{Ibrahim2015Review}.
\C{Immediately following proper attachment of the final kinetochore to the}
 microtubules, inhibitors rapidly dissolve, \C{ultimately} resulting \C{in} active APC/C
(cf. Fig. \ref{fig:sacnetwork}, R9).
\C{APC/C that has been activated} by Cdc20 contributes to the degradation of Cyclin B, \C{which is an} essential requirement for mitotic exit
\cite{Ibrahim2015Review} (cf. Fig. \ref{fig:sacnetwork}, R10).

Simultaneously, APC/C:Cdc20 tags Securin for degradation by proteasome. Securin binds and
inhibits Separase, a protease required to cleave Cohesin,
which is the 'glue' connecting the two sister-chromatids of \C{each} chromosome (cf. Fig. \ref{fig:sacnetwork}, R11).
Thus, activation of APC/C by Cdc20 initiates
sister-chromatid separation, which marks the transition to mitotic exit.
The APC/C activation mechanism is known as SAC silencing, \REV{\C{in which} APC/C itself plays a role, driven by kinetochores, \C{in} the disassembly of
its own inhibitors \cite{varetti2011,Uzunova2012}.} Additionally, many proteins are also involved in the silencing process, \C{such as}
P31$^{\text{comet}}$, UbcH10, and Dynein (\C{see the review} \cite{Ibrahim2015Review}).

Both spindle assembly checkpoint activation and silencing are complex and \C{it is difficult to observe their spatial features experimentally}.
Each SAC component \C{has} various localizations and states upon which the interactions depend.
And a single small human \REV{kinetochore (radius $\sim 0.1 \mu$m)}
has to inhibit all APC/Cs \C{in} the cell \REV{(average radius $\sim 10 \mu$m)},
and after the last attachment, the inhibitors have to be switched off \C{rapidly}.
%
% JAL - I do not understand the meaning of the following sentence well enough to correct the language.
\REV{Also, kinetochores have dynamical complex structure which change over the cell-cycle phases}.
\C{Mitotic} regulation is also challenging theoretically \C{as} computational methods can be hindered by the combinatorial explosion \C{in the number} of
intermediate components (complexes) and explicit representations.
Also, the different components \C{often interact nonlinearly in space and time, and
in the presence of various feedback loops}, which lead to remarkable phenomena that are difficult to predict \cite{Gruenert2010,ibrahim2013,Kreyssig2012,Gruenert2013,Gruenert2015,Sergej2013}.
%
% JAL - 'of the SAC controls' is a clunky phrase - do you mean 'of the SAC' or 'controlling the SAC'?
\C{Mathematical models have helped to illuminate fundamental modules of the SAC controls}
\cite{Doncic2005,Sear2006,Ibrahim2015Cdc20,Lohel2009,Ibrahim2015SACSpatial,Ibrahim2015Pathways,Ibrahim2014convection,Mistry2008,Kreyssig2014,Ibrahim2007,Dennis2013,Ibrahim2008APC,Ibrahim2008Mad2,Ibrahim2009MCC,silencingNatCumm}.
%
%These models \C{have not} yet \C{considered} SAC activation
%\C{at a detailed molecular level in order to distinguish between different pathways.}
These models yet conceived SAC activation
on details molecular level \REV{in order to} distinguish between different pathways.
\REV{However}, none of these models \C{have} considered SAC silencing,
\C{included important properties such as feedback loops \REV{(e.g. the APC/C positive feedback activation loop)}
or have been subjected to stability analyses.}
In this study, a concise model for both SAC activation and silencing in humans was engineered, \C{accounting} for all kinetochore signals
\REV{and also the \C{APC/C} activation loop}.
\C{Ordinary} differential equation (ODE) model simulations \C{were computed}, \REV{along} with a single parameter bifurcation analysis.
\C{The effects of important parameters on the dynamics of the system were then studied, followed by a partial differential equation (PDE) model and its linear stability with various parameter values.}

%%%------------------------------------------------------
%\newpage
\section{Mathematical Framework for the SAC Mechanism}
\subsection{Biochemical Reaction Equations}
The SAC model shown in Fig. \ref{fig:sacnetwork} was \C{manually reduced to a core model} that contains
three biochemical reaction equations (Eqs. (\ref{eq:Mad2}-\ref{eq:stripingMAD2})) \C{describing} the dynamics of the following four species:
I (\REV{Inhibitor production} like Mad2 or Cdc20 or BubR1), I* (cf. MCC), \REV{I*:APC/C (or MCC:APC/C),} and APC/C.
\REV{\C{Four} additional additional reactions} (Eqs. (\ref{eq:CycB_production}-\ref{eq:securin_degradation}))
were \C{incorporated to represent the}
production and degradation of both Cyclin B and Securin.
Mathematically, these additional \REV{reactions} do not affect the \C{overall} dynamics.
The full biochemical reaction rules \C{describing} the SAC system are:
\begin{align}
 &\text{ [I] \hspace{1.75cm}}       \xrightharpoonup{\hspace{.65cm}k_{1},~\text{Un.-Kin\hspace{.7cm}}}\text{[I*]}           \label{eq:Mad2}              \\
 &\text{[I*]} + \text{[APC/C]}  \xrightleftharpoons[k_{-2}]{\hspace{1.3cm} k_2 \hspace{1.2cm}} \text{[I*:APC/C]}        \label{eq:MCCAPC}            \\
 &\text{[I*:APC/C]}\hspace{.5cm}\xrightharpoonup{k_{3},~\text{Att.-Kin},~\text{[APC/C]}}  \text{[I]} + \text{[APC/C]}   \label{eq:stripingMAD2}      \\
 &\text{ \hspace{2.2cm}}        \xrightharpoonup{\hspace{1.25cm}k_{4}\hspace{1.25cm}} 	       \text{[Cyclin B]}        \label{eq:CycB_production}   \\
 &\text{ \hspace{2.2cm}}        \xrightharpoonup{\hspace{1.25cm}k_{5}\hspace{1.25cm}} 	       \text{[Securin]}         \label{eq:Securin_production}\\
 &\text{[Cyclin B] \hspace{.65cm}} \xrightharpoonup{\hspace{.6cm}k_{6},~\text{[APC/C]}\hspace{.6cm}}                    \label{eq:CycB_degradation}  \\
  &\text{[Securin]\hspace{1.cm}}\xrightharpoonup{\hspace{.6cm}k_{7},~\text{[APC/C]}\hspace{.6cm}}                       \label{eq:securin_degradation}
\end{align}
%
%%%------------------------------------------------------

% JAL - You might want to replace 'R3' with Eq. 3
\C{Where Att.-Kin and Un.-Kin refer to attached and unattached kinetochores, respectively.}
\REV{The APC/C positive feedback-loop that is driven by the kinetochore is incorporated in Eq. (\ref{eq:stripingMAD2}.}
\subsection{Ordinary Differential Equation Model}
\label{odemodel}
The reaction rules (Eqs. (\ref{eq:Mad2}-\ref{eq:securin_degradation}))
\C{can be} translated into sets of time-dependent
nonlinear ordinary differential equations (ODEs).
\C{The translation is done by applying the general principles of mass-action kinetics,
computing $dS/dt = Nv(S)$
with state vector $S$, flux vector $v(S)$ and stoichiometric matrix $N$.}
\C{This results in one time-dependent ODE for each species such as Cyclin B or Securin.}
For example the \C{equation} for Cyclin B would be: \REV{${\text{d[Cyclin B]}}/{\text{dt}}={k_4}-{k}_{6}[\text{APC/C}][\text{Cyclin B}]$}.

\subsection{Partial Differential Equations Model }
\label{pdemodel}
Adding a diffusion term \C{to} each differential equation transforms the system \C{into a set of}
coupled partial differential equations (PDEs) \C{known as the \emph{Reaction-Diffusion system},
that are functions of space and time and the following general form:}
\begin{eqnarray}%*
\frac{\partial {[C_{i}]}}{\partial t} =\underbrace{\;
{\text{D}_{i}}\nabla^2
{[C_{i}]\;}}_{\text{Diffusion}}+\underbrace{\;R_{j}\left(\{[C_{i}]\};
\text P\right)\;}_{\text{Reaction}}.
\label{eq:RDS}
\end{eqnarray} % \boldsymbol
Where ${[C_{i}]}$ denote \C{concentrations for species} $i = \{1,...,6\}$.
On the right hand side, the first term refers to the diffusion and the second term represents the biochemical reactions
$R_j = \{R_1,...,R_7\}$ for \C{species} $i$.
The constant coefficient ${\text{D}_{i}}$ represents its species diffusion.
\emph{t} is time and $P$ represents phenomenological parameters.
The operator $\nabla$ refers to the spatial gradient in spherical coordinates
( $\vec{\nabla}(r, \theta,\varphi) = \frac{\partial}{\partial r}\vec{e_r} + \frac{1}{r}\frac{\partial}{\partial \theta}\vec{e_\theta} + \frac{1}{rsin(\theta)}\frac{\partial}{\partial \varphi}\vec{e_\varphi}$).
Recent mathematical models \C{of the} SAC mechanism show that spherical symmetry \C{can be used} without loss of generality  \cite{Ibrahim2014convection,Ibrahim2015SACSpatial}, \C{and this form is used here}.
The system in Eq. (\ref{eq:RDS}) \C{reduces} to the following PDEs depending on $t$ and $r$:
\begin{eqnarray}%*
\frac{\partial {[C_{i}]}}{\partial t} =
\underbrace{\; \frac{D}{r^2} \frac{\partial}{\partial r} (r^2 \frac{\partial {[C_{i}]}}{\partial r})   }_{\text{Diffusion}}+
\underbrace{\;R_{j}\left(\{[C_{i}]\};\text P\right)\;}_{\text{Reaction}}.
\label{eq:RDSsymm}
\end{eqnarray} % \boldsymbol
\subsection{Numerical Simulations}
\label{numerical}
The \C{ODE models were implemented} in the \C{freely-available} software package XPPAUT \cite{xppaut2002} and integrated using \C{a stiff solver}.
The bifurcation \C{analyses} were computed with AUTO \cite{auto1981} using the XPPAUT interface.
The \C{PDE} reaction-diffusion systems were implemented in matlab
(MathWorks) and simulated using the pdepe function.
The pdepe solver can \C{handle} systems of parabolic and elliptic PDEs in one space variable and time, \C{perfectly matching the presented SAC model.}
This function uses the method of lines, \C{spatially discretizing the problem in space and converting it to}
a system of ordinary differential equations
that can \C{be solved} using the numerical stiff solver ode15s in Matlab \cite{pdepe1990}.
The ode15s ODEs solver \C{can} solve \C{the} differential-algebraic systems that frequently arise in PDE systems.

\REV{The same initial concentrations were applied to all models for comparability and consistency (Table~\ref{tab:kinetics}).
The \C{specific values were} chosen according to data from the literature (\cite{Tang2001,stegmeier2007anaphase,Fang2002,Howell2000,Mad2Concentration,Moore2003,cyclinB2,cyclinB,separase}).
The kinetic rate constants \C{were} also taken from literature \C{when available} (e.g. \cite{cyclinbSecurin,cyclinbSecurin2,DeAntoni2005,Mapelli2006,Ibrahim2008APC};
Table~\ref{tab:kinetics}).
For the rate constants {($k_1$, and $k_3$, see Table~\ref{tab:kinetics})},
several computations with values representing the whole physiologically reasonable parameter range
were compared (see Fig. \ref{fig:ode}D and Fig. \ref{fig:parameters}).
% and $\alpha$
The trajectories are discussed in corresponding figures.
% JAL - This last sentence is odd - which figures are you referring to specifically and what do they correspond to?
}

\subsection{Model Assumptions}
\label{assumptions}
\C{In the PDE simulations, the cell is taken} to be a sphere with radius $R$. The last unattached kinetochore is modeled as \C{a sub-sphere} (radius $r$) located in the center of the cell.
Boundary conditions are assumed to be reflective and the numbers of all interacting elements \C{are assumed to be
conserved. The PDEs} are assumed \REV{to be} spherically symmetric.
\REV{Reactions are considered to follow the mass-action-kinetics law and all rate
constants are listed in (Table~\ref{tab:kinetics}).
% JAL - I'm unclear what the sentences below mean - you say that various things are localized in different places
% but that uniform distribution is the most appropriate assumption?
The initial conditions for the PDEs are \C{correspond to uniform distribution.}
The reason is that APC/C \C{is} found experimentally to be localized at kinetochores, spindles and poles
\cite{APC_localization1,APC_localization2} \C{while MCC is found to be localized
in both the kinetochores and the} cytosol \cite{Ibrahim2009MCC}.
Thus \C{it is most appropriate to assume a uniform distribution.}
}

\newpage
\section{Results and Discussions}

\subsection{Ordinary Differential Equation SAC Model}

\begin{figure}[htp]
\begin{center}
\includegraphics[width=1\textwidth]{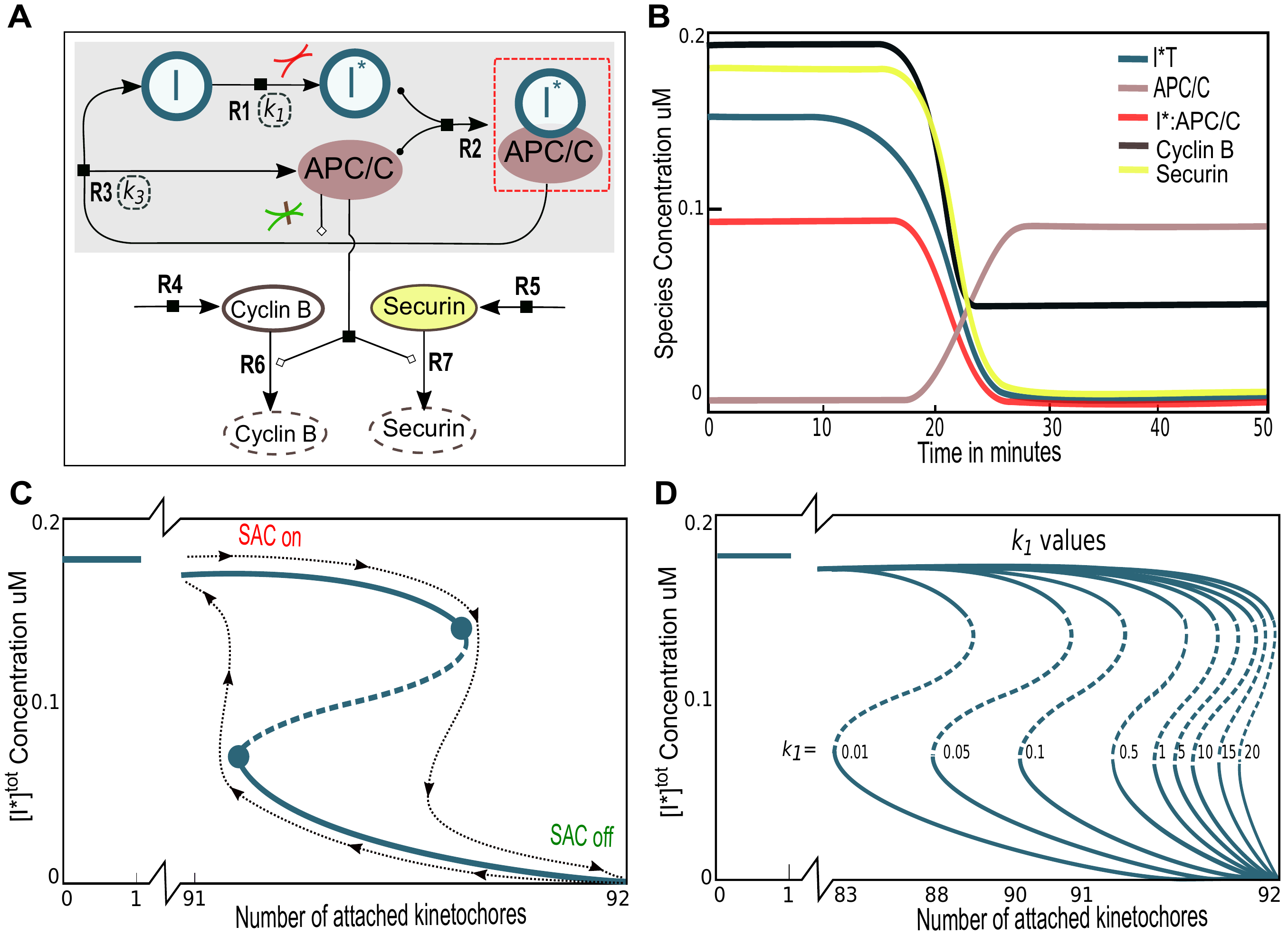}
\caption{Spindle Assembly Checkpoint ODE model.
    (A) Biochemical reaction network \C{for} SAC activation and silencing.
    I is the inhibitor producing, for example, \C{BubR1, Mad2 and Cdc20}.
    I* is a potent APC/C inhibitor which mainly refers to \C{the} MCC complex in this model.
    \C{Production of I*} is enhanced by \C{signals from} unattached kinetochores.
    \C{This is SAC activation (or APC/C inhibition)}.
    \C{Directly following attachment of the final kinetochore to the spindle microtubules},
    the inhibitor dissolves and APC/C is activated.
    \C{This is SAC silencing (or APC/C activation).
    \REV{A positive feedback loop takes place via APC/C itself, which is driven by kinetochores}.}
    Nodes represent core SAC proteins or complexes, edges refer to their interactions.
    (B) Numerical solution of the \C{ODE} model \C{showing the concentration of each SAC component versus time.}
    Once all kinetochores are attached (at 17 minutes), APC/C is activated. Active APC/C tags both Securin and Cyclin B for degradation.
    Cyclin B is \C{degraded in two phases, first by APC/C:Cdc20 and second via APC/C:Cdh1 during mitotic exit.}
    Thus Cyclin B level does not reach zero in our simulation, which is consistent with \C{the} literature \cite{Irniger2002}.
    All parameter \C{values are set} according to Table~\ref{tab:kinetics} (see the text for more details).
    (C) \C{Single parameter bifurcation curve showing kinetochore signals versus total I*.}
    Unstable saddle points are shown by dashed lines and stable node points \C{by solid lines}.
    Both stable and unstable states meet at saddle node bifurcation points shown by solid circles.
    The SAC checkpoint \C{is released and APC/C activated only when almost all kinetochores} are attached (approximately 91.98).
    As the cell enters anaphase, I* \C{falls} back to zero.
    \C{The black line indicates how the switch flips from the SAC-active state to the SAC-inactive state as number of attached kinetochores increases.}
    (D)The bifurcation curves are sensitive to $k_1$ values (see text for details).
%
%    k3 value (must be between 0.5 to 0.005).
%    with 0.5  is  91.85
%    with 0.1  is  91.98
%    with 0.05 is  91.99
%    with 0.005is  91.999
%    while ka (all before with ka=10) as follows
%    with ka=1 is 91.80
%    with ka=10 is 91.98
%    with ka=100 is 91.998
%    %
%    max shift to left reached  by
%    ka=1.1  and kcat to 0.5 leads to a point at 90.73
}
  \label{fig:ode}
\end{center}
\end{figure}

% JAL - Perhaps 'Inhibitors producing' -> 'Inhibition' or 'Inhibitor production'? I will leave unchanged for now.
\REV{The SAC mechanism} (Fig. \ref{fig:sacnetwork}) consists of four modules:
\REV{Inhibitors production} (Fig. \ref{fig:sacnetwork} red box), \REV{APC/C} inhibition
(Fig. \ref{fig:sacnetwork} yellow box), APC/C activation (Fig. \ref{fig:sacnetwork} green box),
and mitotic exit (Fig. \ref{fig:sacnetwork} blue box).
\REV{This mechanism} was reduced to \REV{a minimal core model} (shaded area shown in Fig. \ref{fig:ode}A)
\C{comprising} three reaction equations (R\ref{eq:Mad2}-R\ref{eq:stripingMAD2}) \C{that} describe the dynamics of four main species:
I (represents either Mad2 or Cdc20 or BubR1), I* (cf. MCC), \REV{I*:APC/C (or MCC:APC/C),} and APC/C.
These reactions were translated into a set of coupled nonlinear ordinary differential \C{equations}
under the assumption of mass action kinetics for all reactions.
In order to \C{further reduce the model}, the following biochemical assumptions were assumed.
The total concentration of APC/C \C{in the system is constant} and can be \C{calculated using}
$\text{[APC/C]}^{tot} = [APC/C] + [I*:APC/C]$.
The same is true \C{for $I$}, using $\text{[I]}^{tot}$ = [I] + [I*] + [I*:APC/C].
For simplicity, the total amount of I* was \C{taken to be} $\text{[I*]}^{tot}$=[I*]+[I*:APC/C].
\C{The core SAC system can be written as the following pair of nonlinear ODEs:}
\begin{equation}
\frac{\text{d[I*]}^{tot}}{\text{d}t} = k_1. U(\text{[I]}^{tot}-\text{[I*]}^{tot})-k_{3}.A\text{\REV{[APC/C]}[I*:APC/C]})
\label{eq:conmcc}
\end{equation}
\begin{equation}
\frac{\text{d[APC/C]}}{\text{d}t} = -k_2\text{[I*][APC/C]}+ (k_{-2} + A\text{\REV{[APC/C]}}) \text{[I*:APC/C]  }
\label{eq:conapc}
\end{equation}
Finally, \REV{an extended model is generated by adding Cyclin B, Securin, and also presuming}
that APC/C \C{is} in steady state, \C{so that} putting the total concentration of {[APC/C]$^{tot}$} into the system
\C{gives} the following full \REV{ODE-based model}:
\REV{
\begin{align}
\frac{\text{d[I*]}^{tot}}{\text{d}t} &= k_1. U(\text{[I]}^{tot}-\text{[I*]}^{tot})-k_{3}.A(\text{[APC/C]}^{tot}-\text{[I*:APC/C]})\text{[I*:APC/C]})
\label{eq:conmcc_b}\\
\text{[I*:APC/C]} &= \frac{-b \pm \sqrt{b^2 -4ac}}{2a}
\label{eq:conapc_b}\\
\frac{\text{d[Cyclin B]}^{tot}}{\text{d}t} &= {k_4}-{k}_{6}[\text{APC/C}][\text{Cyclin B}]
\label{eq:cyclinB_b}\\
\frac{\text{d[Securin]}^{tot}}{\text{d}t} &= {k_5}-{k}_{7}[\text{APC/C}][\text{Securin}]
\label{eq:securin_b}
\\\nonumber
\text{where,}\hspace{1.7cm}&                                                          \\\nonumber
a &= -k_2 - A                                                                   \\\nonumber
b &= k_2\text{[I*]} + k_2\text{[APC/C]}^{tot} + k_{-2} + A.\text{[APC/C]}^{tot} \\\nonumber
c &= -k2 \text{[I*]}\text{[APC/C]}^{tot}                                        \nonumber
\end{align}}
\REV{\C{And} $U$ refers to the number of unattached kinetochores \C{that have} an additional ODE equation:
$\text{d}{U}/\text{dt} = -\alpha.{U}$.
\C{$A$ refers} to the number of attached kinetochores and \C{is equal to} $A=92-U$. }
\REV{It is clear that the addition of Cyclin B and/or Securin \C{does} not affect the system's behavior
\C{because these species depend on the} active APC/C level.}

The \C{results of} numerical \REV{simulations (see Section \ref{numerical})} \C{are} shown in Fig. \ref{fig:ode}B.
\REV{
\C{They are consistent with experimental findings in terms} of APC/C level and \C{the timing of anaphase} \cite{Bakhoum2009,Lohel2009}.
The concentration of APC/C (Fig. \ref{fig:ode}B brown line) is very low even when a single kinetochore is not attached.
After around 18 minutes, the APC/C activity increases rapidly to reach its maximum value.
The complexes I*:APC/C (or MCC:APC/C) \C{exhibit behavior that is opposite to that of} APC/C (Fig. \ref{fig:ode}B, red line).
I* concentration \C{follows I*:APC/C with a difference only} in the initial concentration (Fig. \ref{fig:ode}B, blue line).
}

\C{A} \REV{one parameter bifurcation analysis} was performed for the nonlinear ODE-based system
(Eqs. \ref{eq:conmcc_b} and \ref{eq:conapc_b}).
\C{Simulations of the bifurcation curves were performed using the} AUTO software (Section \ref{numerical}).
The \C{desired behavior is a bistable switch} influencing \C{the} total I* \C{as kinetochores are} attached one-by-one.
Fig. \ref{fig:ode}C shows the \C{result, which is} a typical S-shape in the number of attached kinetochores \C{as a function of the} total concentration of the inhibitor, I*.
Solid lines \C{refer to stable nodes while dashed} lines refers to the unstable saddles.
Stable and unstable states meet at saddle node bifurcation points that \C{are indicated} by solid circles.
When nearly all kinetochores are attached (91.85 kinetochores), the SAC checkpoint \C{turns} off and APC/C is \C{activated
rapidly}.
\C{Total I* falls back to zero as the cell exits} from mitosis.
\C{The black line indicates how the switch alternates between the SAC-active and SAC-inactive states as the
number of attached kinetochores increases.}
All parameters were taken from Table~\ref{tab:kinetics}.

\begin{figure}[htp]
\begin{center}
\includegraphics[width=.6\textwidth]{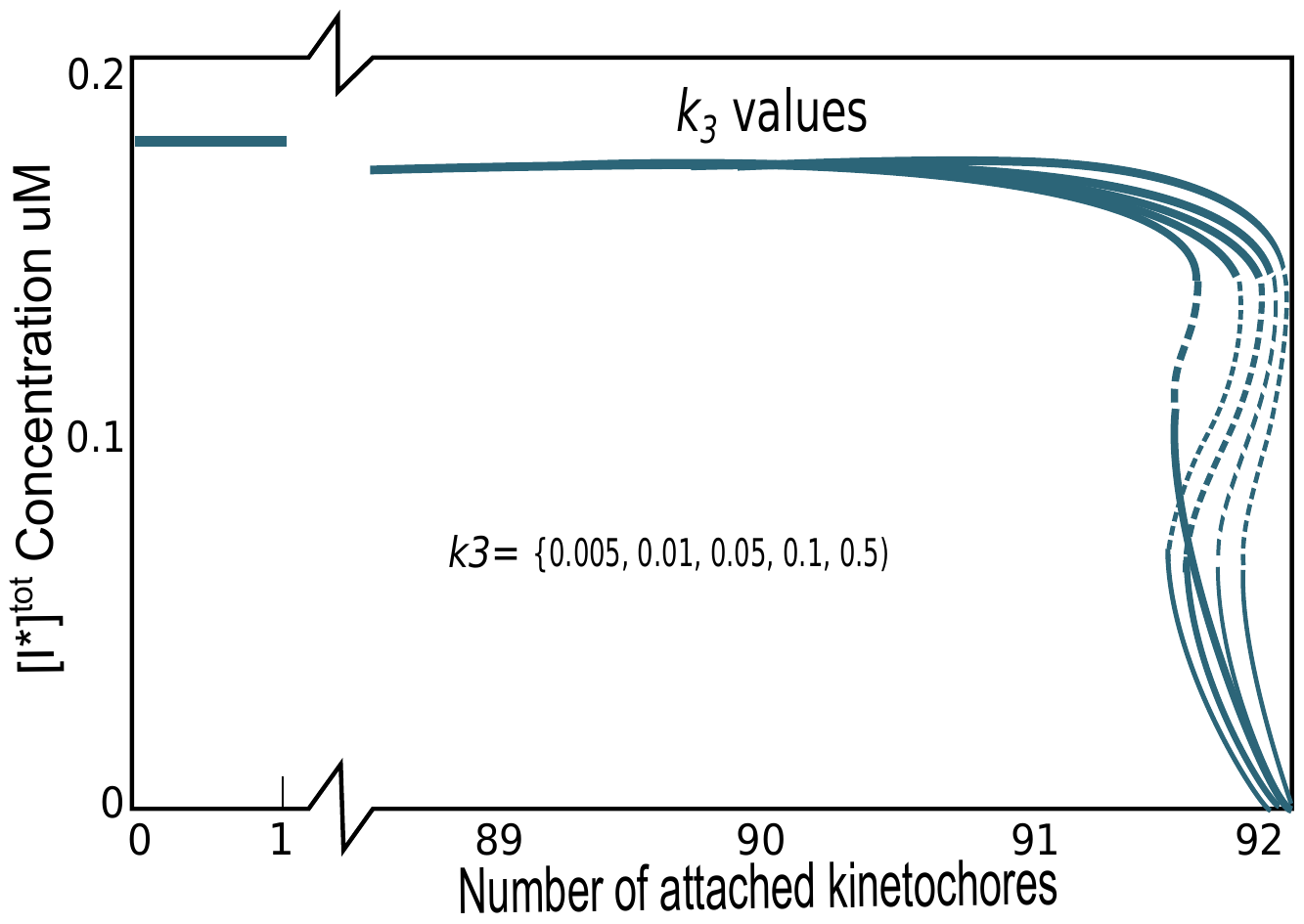}
\end{center}
\caption{Sensitivity to $k_3$ values.
    Bifurcation curve for the unattached kinetochore signals versus the total $I*$ level.
    Different curves shown the sensitivity of \C{the analysis to $k_3$ values}.
    Unstable saddle points are shown by dashed lines and stable nodes points \C{by solid lines}.
    Both stable and unstable states meet at saddle-node bifurcation points shown by solid circles.
    }
    \label{fig:parameters}
\end{figure}

In order to determine the sensitivity \C{to} parameter values, \C{various ranges of parameter values were examined}
(Fig. \ref{fig:ode}D and Fig. \ref{fig:parameters}).
The bifurcation diagram is most sensitive to $k_1$ values. The working range is between \C{0.01$\pers$ and 30$\pers$}.
As $k_1$ increases, the bifurcation point and \C{curve are shifted} to the right (Fig. \ref{fig:ode}D).
Low values of $k_1$ (c.f. 0.01$\pers$) \C{initiate the SAC switch before 83 kinetochores are attached, which could cause grave risk to the cell.}
All other $k_1$ values \C{higher} than 0.5$\pers$ {generate a safe and appropriately-timed switch.}
$k_3$ values have some minor effects on the system (Fig. \ref{fig:parameters}). Its working values are between \C{0.01$\pers$ ad 100$\pers$}.
Increasing $k_3$ value to 100 \C{can} shift the bifurcation point higher to meet at higher amount of \C{total I*}.
%    k3 is kcat value (must be between 0.5 to 0.005).
%    with 0.5  is  91.85
%    with 0.1  is  91.98
%    with 0.05 is  91.99
%    with 0.005is  91.999
%
%    while k1 is ka (all before with ka=10) as follows
%    with ka=1 is 91.80
%    with ka=10 is 91.98
%    with ka=100 is 91.998
%    %
%    max shift to left reached  by
%    ka=1.1  and kcat to 0.5 leads to a point at 90.73
%%%------------------------------------------------------
\clearpage

\subsection{Partial Differential Equation SAC Model}
Several studies \C{have indicated} the importance of diffusion in \C{the} SAC mechanism (e.g. \cite{Doncic2005}, \cite{Ibrahim2015SACSpatial}).
However, these models \C{represented} either smaller budding yeast cells \REV{\C{using PDEs or detailed SAC activation using ODEs}}.
\C{Here, a \REV{minimal core SAC} model represented by PDEs is examined.}

\begin{figure}[htp]
\begin{center}
\includegraphics[width=1.\textwidth]{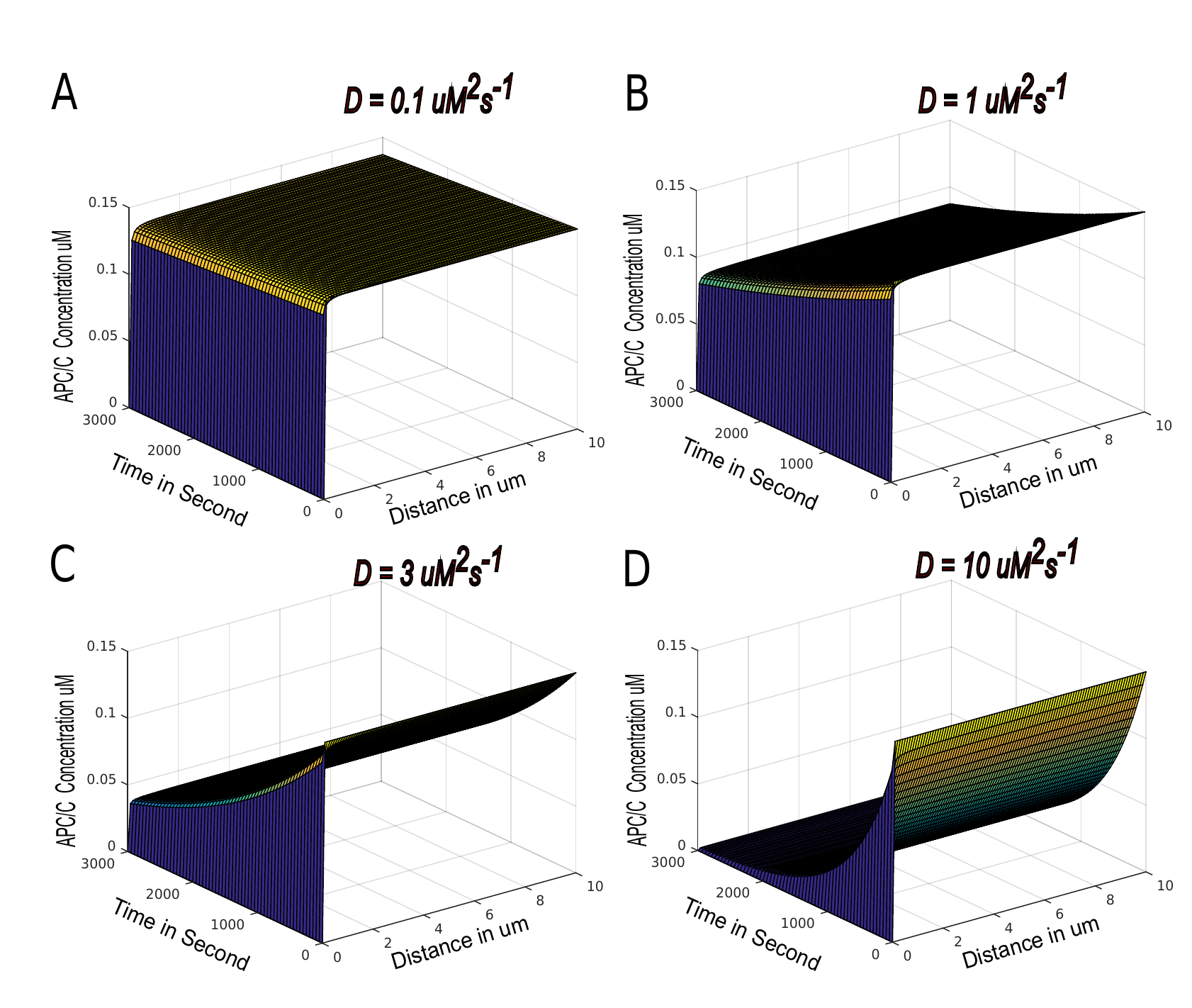}
\end{center}
\caption{Partial differential equation model simulations.
    \REV{
    \C{Simulation diagrams (A-B) are based} on low diffusion coefficient values (0.1${\mu m^2}{s^{-1}}$ and 1${\mu m^2}{s^{-1}}$), while (C-D) are based on realistic and higher values (3${\mu m^2}{s^{-1}}$ and 10${\mu m^2}{s^{-1}}$).}
    }
    \label{fig:pde}
\end{figure}

\REV{Cyclin B and Securin reactions are eliminated \C{because} these reactions are \C{the} output of active APC/C
and cannot affect APC/C regulation}.
To this end, a second derivative diffusion term was added to the SAC system (Eq.\ref{eq:conapc}-Eq.\ref{eq:conmcc}),
\C{which leads to a set of coupled partial differential equations known as a}
\emph{Reaction-Diffusion system} with spherical symmetry (Section \ref{pdemodel}):
\begin{equation}
\frac{\text{d[I*]}^{tot}}{\text{d}t} = \frac{D}{r^2} \frac{\partial}{\partial r} (r^2 \frac{\partial {\text{[I*]}^{tot}}}{\partial r})
                                    +k_1. U(\text{[I]}^{tot}-\text{[I*]}^{tot})-k_{3}.A\text{[APC/C][I*:APC/C]})
\label{eq:mccpde}
\end{equation}
\begin{equation}
\frac{\text{d[APC/C]}}{\text{d}t} = \frac{D}{r^2} \frac{\partial}{\partial r} (r^2 \frac{\partial {[APC/C]}}{\partial r})
                                    -k_2\text{[I*][APC/C]}+ (k_{-2} + k_{3}.A\text{[APC/C]}) \text{[I*:APC/C]  }
\label{eq:apcpde}
\end{equation}
\C{These are supplemented by reflective (Neumann) boundary conditions and assumptions of spherical geometry for the cell and kinetochores (Section \ref{assumptions}).}
\REV{Numerical simulations of the PDE system (Eq.\ref{eq:mccpde}-Eq.\ref{eq:apcpde}) were conducted in matlab.
The diffusion coefficient for APC/C is known from the literature (Table~\ref{tab:kinetics}) while \C{that of} MCC (or $I*$)
is not. Therefore the simulations \C{were} repeated using \C{four different values for the MCC diffusion coefficient},
(Table~\ref{tab:kinetics} and Fig. \ref{fig:pde}A-D).
One of these values \C{(referred to here as `realistic') is chosen with reference to experimental work on MCC subunits
(cf. Mad2, BubR1, Bub2, and Cdc20; see Table~\ref{tab:kinetics}).
The other values span a wide range to allow any effects of diffusion to be observed.}
Following the idea by Doncic et. al \cite{Doncic2005}, only the last unattached kinetochore \C{was} considered.
Hence, the parameter $U$ \C{was} assumed to be constant \C{and to refer only to the last} unattached
kinetochore (cf. $U=1$, and $A=92-U=91$).
The aim \C{was} to reveal, if \C{the} last kinetochore is not yet attached, how fast MCC is able to inhibit \C{high levels} of APC/C in the cell.

The simulations show that using \C{a realistic diffusion coefficient} for MCC (3${\mu m^2}{s^{-1}}$),
APC/C is inhibited very slowly after an hour (Fig. \ref{fig:pde}C).
Using lower diffusion coefficients (1${\mu m^2}{s^{-1}}$), only 10\% of \C{the} APC/C level is inhibited (Fig. \ref{fig:pde}B) while
APC/C is not inhibited at all (Fig. \ref{fig:pde}A) \C{when} using an even lower value (0.1${\mu m^2}{s^{-1}}$) .
\C{Using a diffusion coefficient higher than the realistic value (10${\mu m^2}{s^{-1}}$) achieves proper inhibition of APC/C (Fig. \ref{fig:pde}D).}
%
%These results is sensitive to the combination with reaction rate constants.
%
\C{The conclusions are that a realistic diffusion coefficient is sufficient for inhibiting APC/C, albeit slowly, but a coefficient 2-3 times larger is required for proper SAC functioning.}
\C{Thus, it is recommended to use a SAC model based on a PDE model, rather than solely on ODEs, though it is not essential.}
}
\\[2cm]

%
%%%------------------------------------------------------

\newpage
\subsection{Linear Stability Analysis}

As shown in the previous section, \C{a realistic value for the diffusion coefficient does not have a strong effect} on the SAC model.
However, this does not exclude the possibility that diffusion \C{influences properties of the system such as its stability.}
\C{To study this, a} linear stability analysis \C{was performed on the SAC PDE} model (Eqs.\ref{eq:mccpde}-\ref{eq:apcpde}).

\REV{\C{Taking} $\text{[APC/C]}^{tot}$ = [APC/C] + [I*:APC/C], and
$\text{[I*]}^{tot}$=[I*]+[I*:APC/C] (or equivalently [I*]=$\text{[I*]}^{tot}-\text{[APC/C]}^{tot}$- [APC/C]), the system will be:

\begin{equation}
\frac{\text{d[I*]}^{tot}}{\text{d}t} = \frac{D}{r^2} \frac{\partial}{\partial r} (r^2 \frac{\partial {\text{[I*]}^{tot}}}{\partial r})
                                    +k_1. U(\text{[I]}^{tot}-\text{[I*]}^{tot})-k_{3}.A\text{[APC/C]}(\text{[APC/C]}^{tot} - \text{[APC/C]})
\label{eq:mccpde_n}
\end{equation}
\begin{equation}
\begin{split}
\frac{\text{d[APC/C]}}{\text{d}t} = &\frac{D}{r^2} \frac{\partial}{\partial r} (r^2 \frac{\partial {[APC/C]}}{\partial r})
                                    -k_2(\text{[I*]}^{tot}-\text{[APC/C]}^{tot}- [APC/C])\text{[APC/C]}+\\
                                    & (k_{-2} + k_{3}.A\text{[APC/C]})(\text{[APC/C]}^{tot} - \text{[APC/C]})
\label{eq:apcpde_n}
\end{split}
\end{equation}

\C{Assuming} small perturbations \C{on} I* and APC/C in standard form (see for example \cite{logan1997}):

\begin{equation}
\text{[I*]}^{tot}(r,t) = \text{[I*]}^{tot}_1(r) + \text{[I*]}^{tot}_2(r,t)\label{eq:m1m2}\\
\end{equation}

\begin{equation}
\text{[APC/C]}(r,t) = \text{[APC/C]}_1(r) + \text{[APC/C]}_2(r,t)\label{eq:a1a2}\\[.5cm]
\end{equation}
where $\text{[I*]}_1$ and $\text{[APC/C]}_1$ denote the steady state, \C{and}
$\text{[I*]}_2$ and $\text{[APC/C]}_2$ denote the unsteady-state (or disturbance).
Substituting Eqs.\ref{eq:m1m2}-\ref{eq:a1a2} into \C{the PDE} system (Eqs.\ref{eq:mccpde}-\ref{eq:apcpde})
\C{leads to two separate systems}.
The linearized equations governing the disturbance system \C{are}

\begin{equation}
\begin{split}
\frac{\text{d}[\text{I*}]^{tot}_2}{\text{d}t} =& \frac{D}{r^2} \frac{\partial}{\partial r} (r^2 \frac{\partial {[\text{I*}]}^{tot}_2}{\partial r})
                                    - \beta_1([\text{I*}]^{tot}_2)-
                                    k_{3}.A(\text{[APC/C]}_2\text{[APC/C]}^{tot}+ \\
                                    & 2\text{[APC/C]}_1\text{[APC/C]}_2 + \text{[APC/C]}_2^2)
\label{eq:mccstability2}\\
\end{split}
\end{equation}
\begin{equation}
\begin{split}
\frac{\text{d[APC/C]}_2}{\text{d}t} =& \frac{D}{r^2} \frac{\partial}{\partial r} (r^2 \frac{\partial \text{[APC/C]}_2}{\partial r})
                                     + (\beta_2\text{[APC/C]}_2  -k_2\text{[I*]}^{tot})\text{[APC/C]}_2 +\beta_3 - \\
                                     &(k_2 +k_{3}.A)\text{[APC/C]}^2_2
\label{eq:apcstability2}
\end{split}
\end{equation}

}
\C{where, $\beta_1 =k_1.U$, $\text{[APC/C]}^{tot}$ is constant,
$\beta_2 = (k_{2}\text[I*]^{tot}_2 + (k_2+k_{3}.A)\text{[APC/C]}^{tot})$, and $\beta_3 = k_{-2}\text{[APC/C]}^{tot}$.}
Note that for hydrodynamic stability any second order terms in the disturbance system should be neglected \cite{logan1997}.
Since the diffusion coefficient for I* (MCC complex) is unknown, the focus \C{is on} I*.

\C{The solution is assumed to be a propagating wave, dependent on both space and time,
and given by \REV{$[I*]^{tot} = I_0 e^{i\gamma{r-ct}}$,
and $[APC/C] = \REV{APC/C_0} e^{i\gamma{r-ct}}$}}.
\C{\REV{$\gamma$} is the wave number and $C$ is the complex speed of the wave $C = C_r + iC_i$
where $C_i$ is the important parameter for stability analysis.} \C{The} disturbance is damped
if $C_i$ is \C{negative, whereas} for $C_i>0$, the disturbance will grow and \C{lead to instability.
After simplification, the relation becomes:}

$ C_i = -k_1.U -  k_{3}.A* (\REV{(APC/C_0)} / I_0) - D. \REV{\gamma}^2 < 0$ always.

This indicates that the system is stable \C{regardless of the value of the diffusion coefficient.}
\C{The diffusion coefficient plays no major in the SAC model presented here because all kinetic reaction constants are already high.}

\section{Conclusion}
\REV{
The \REVV{spindle} assembly checkpoint (SAC) regulates the timing of chromosome segregation to prevent the creation of aneuploidy
and tumorigenesis \cite{Morais2013,Holland2009,denisenko2016}.
SAC is highly sensitive to kinetochore–-microtubule attachment, \C{such that} even a single unattached or mis-attached \C{kinetochore
can maintain checkpoint activation and keep the} APC/C complex inactive.
\C{The checkpoint is only silenced when all chromosomes are properly attached to microtubules, after which APC/C is activated and anaphase onset occurs.}
The mechanism by which SAC activation and silencing is achieved is still largely elusive.

Despite several systems-level mathematical efforts \C{in recent years to model SAC activation,
no model has been built with the level of detail in its mathematical abstraction that is crucial} for further developments \cite{Ibrahim2015Review}.
For example, no model \C{has considered all kinetochore signals. In addition}, to the best of our knowledge, there is no model
\C{of the SAC that combines both SAC activation and silence, or one that includes any kind of feedback.}
\C{This is the first work presenting a mathematical analysis that integrates both SAC activation and silencing.}
The model is minimal \C{in that it} contains \C{only} four reactions and four species, considers all 92 kinetochore signals,
and incorporates \C{kinetochore-driven feedback} on APC/C activation (for SAC silencing).
The model \C{incorporates differing levels of detail}. The ODE-based model \C{was simulated and generated}
a bifurcation diagram realistic \C{switch behavior over a} wide range of parameters.
The model was also compared with \C{experimental findings and found to be consistent} \REVV{(Table~\ref{tab:mutation})}.

\C{Parameters were selected carefully.}
All models have the same initial concentrations which \C{were} chosen according to data from the literature \REVV{(Table~\ref{tab:kinetics})}.
The reaction rate constants \C{were} also taken from literature \C{where they were known}. \C{For two} rate constants $k_1$, and $k_3$,
several computations \C{were performed using values representing the whole physiologically reasonable parameter range}
(see Fig. \ref{fig:ode}D and Fig. \ref{fig:parameters}).

\C{It had been thought that diffusion is} required for proper SAC functioning (e.g. \cite{Ibrahim2015Review}).
\C{To investigate the role of diffusion, a second level of detail was implemented in a PDE-based reaction-diffusion model.}
\C{Active transport (or convection) was not considered because there is no experimental evidence
for transport of either APC/C or MCC.}
\C{Active transport is essential for inhibitor formation, which has been studied extensively in the literature and is therefore not considered here.}
\C{It has} been shown both experimentally and mathematically that active transport of O-Mad2 towards the spindle mid-zone increases the efficiency of Mad2-activation \C{in inhibiting} Cdc20 \cite{Ibrahim2014convection,Howell2000,mad2transport}.
The diffusion coefficient for APC/C is known \cite{wang2006}. The MCC diffusion \C{coefficient is not known but was} approximated
based on its sub-components (Mad2, Cdc20, BubR1 and Bub3). \C{This value of} (3~${\mu m^2}{s^{-1}}$) was found to be
insufficient for rapid APC/C inhibition. \C{Fast APC/C inhibition, and therefore proper SAC functioning, was found to require a value that was}
at least twice \C{as high}. This may be due to noise in the experimental data or because there is an additional mechanism that increase the rate independent of the kinetochores (\cite{Ibrahim2009MCC}).

The \C{mathematical modeling and analysis presented here can serve as a basis}
for more sophisticated models be used to
evaluate novel hypotheses related to mitosis and \C{the} cell cycle.
\C{A combination of further experimental work and mathematical analysis be necessary
to fill in the gaps in our understanding of the cell life cycle.}
}
%%%------------------------------------------------------
\newpage
\begin{table}[ht]
\caption{\bf{Kinetic \C{parameters of the SAC model}}}
\REV{\begin{tabular}{p{0.09\textwidth}p{0.15\textwidth}p{0.1\textwidth}p{0.12\textwidth}p{0.13\textwidth}p{0.115\textwidth}p{0.105\textwidth}}
\hline\noalign{\smallskip}
            & \bf Symbol                   &\bf Fig. 2B, C&\bf Fig. 2D& \bf Fig. 3 & \bf Fig. 4 & \bf Remark                  \\
            &                              &             &           &            &            &                             \\\hline
%%%------------------------------
\multicolumn{2}{l}{\underline{Initial amount}}         &&&&			   			               		                                 \\[.1cm]
			&APC/C                         &0.09 $\mu M$   &0.09 $\mu M$ &0.09 $\mu M$ & 0.09 $\mu M$ & \cite{Tang2001,stegmeier2007anaphase} \\
			&I*(cf. MCC)                   &0.15 $\mu M$   &0.15 $\mu M$ &0.15 $\mu M$ & 0.15 $\mu M$ & \cite{Ibrahim2015Pathways}   \\
            & I                            &0.15 $\mu M$   &0.15 $\mu M$ &0.15 $\mu M$ &0.15 $\mu M$  & \cite{Fang2002,Howell2000,Mad2Concentration}             \\
            & I*:APC/C                     &0              &0            &0            & 0            &                                         \\
            & Cyclin B                     &0.2 $\mu M$    &0.2 $\mu M$  &0.2 $\mu M$  &0.2 $\mu M$   & \cite{Moore2003,cyclinB2,cyclinB}     \\
            & Securin                      &0.18 $\mu M$   &0.18 $\mu M$ &0.18 $\mu M$ &0.18 $\mu M$  & \cite{separase}                        \\
            & U                            &92             &      92     &        92   &92            & \cite{Ibrahim2008APC,Ibrahim2015Review}     \\
            & A                            &92-U           &      92-U   &        92-U &92-U          & \cite{Ibrahim2008APC,Ibrahim2015Review}  \\
%			&Cdc20:C-Mad2                  &                                                          & \cite{Fang2002, Howell2000}	\\
            %-------------------------------------------------------------------------------------
\hline\\[-.3cm]\noalign{\smallskip}
\multicolumn{2}{l}{\underline{Diffusion \C{coefficients}}} &      &             &             &	                                                       \\[.1cm]
            &APC/C	 	                   &               &             &             & 1.8~${\mu m^2}{s^{-1}}$  & \cite{wang2006}  		       \\
            &I* (cf. MCC)                  &               &             &             & 0.1-10~${\mu m^2}{s^{-1}}$ & This study 		               \\
            &I (Cdc20)	 	               &               &             &             & 19.5~${\mu m^2}{s^{-1}}$ & \cite{wang2006}  		       \\
            &I (Mad2)	 	               &               &             &             & 5~${\mu m^2}{s^{-1}}$    & \cite{Ibrahim2015SACSpatial}   \\
            &I (Mad2:Cdc20)                &               &             &             & 4~${\mu m^2}{s^{-1}}$    & \cite{Ibrahim2015SACSpatial}   \\
            &I (BubR1:Bub3)                &               &             &             &7.9~${\mu m^2}{s^{-1}}$   &\cite{Tang2001,Fang2002}        \\
\hline\\[-.3cm]\noalign{\smallskip}
\multicolumn{2}{l}{\underline{Environment}}   &&&&			                			               	                                              \\[.1cm]
            &Radius of the kinetochore	   &               &             &             &0.1 $\mu m$                  & \cite{cherry1989kinetochoresize}\\
            &Radius of the cell            &               &             &             &10 $\mu m$                  & \cite{Ibrahim2014convection}    \\
\hline\\[-.3cm]\noalign{\smallskip}
\multicolumn{2}{l}{\underline{Rate constants}} &&&&	            		    			               		                                          \\[.1cm]
            & $\alpha$                     &$ 0.0032\pers$ &$0.0032~\pers$&$0.0032~\pers$ & $0.0032~\pers$             &                              \\
            & $k_1$		                   &$ 10~\pers$     &$0.01-20~\pers$&$10 ~\pers$   &  $10 ~\pers$		       & This study 			      \\
            & $k_2$		                   &$ 100~\perM$    &$100~\perM$  &$100~\perM$   & $100~\perM$		           &\cite{Ibrahim2008APC}         \\
            & $k_{-2}$                     &$ 0.08~\pers$   &$0.08 ~\pers$ &$0.08 ~\pers$  & $0.08 ~\pers$		       &\cite{Ibrahim2008APC}         \\
            & $k_{3}$                      &$ 0.001~\pers$  &$ 0.1~\pers$  &$0.005-0.5~\pers$& $ 0.1~\pers$    	       & This study 			      \\
            & $k_{4}$                      &$ 0.011~\pers$  &$0.011 ~\pers$&$ 0.011~\pers$ &  	                       &               \\
            & $k_{5}$                      &$ 0.0019~\pers$ &$0.0019 ~\pers$&$0.0019~\pers$&                           &               \\
            & $k_{6}$                      &$ 0.058~\pers$  &$0.058~\pers$  &$ 0.058~\pers$&                           &\cite{cyclinbSecurin,cyclinbSecurin2} \\
            & $k_{7}$                      &$ 0.0095~\pers$ &$0.0095 ~\pers$&$ 0.0095~\pers$&                          &\cite{cyclinbSecurin,cyclinbSecurin2} \\
%
%
% %$k_{attach}$   & 		            & $0.0032 \pers$ & takes around 20 min				                                    \\
\hline\noalign{\smallskip}
\end{tabular}
\\[.2cm]
{Diffusion for MCC was calculated \C{using} $D_{A.B} = \frac{D_A.D_B}{D_A + D_B}$,
where $D_A$ and $D_B$ are the diffusion coefficients for $A$ (Mad2:Cdc20) and $B$ (BubR1:Bub3), respectively.
}
}
\label{tab:kinetics}
\end{table}

\begin{table}[ht]
\REV{
\caption{\bf{Comparison of the \C{{\bf\it in vitro} and {\bf\it in silico}} mutation experiments with respect to APC/C activity}}
\label{tab:mutation}
\begin{tabular}{p{0.1\textwidth}p{0.13\textwidth}p{0.25\textwidth}p{0.25\textwidth}p{0.15\textwidth}}\hline
%--------------------
\bf Type of mutation&\bf Experiment&\bf{\it in-vitro}                     &\bf{\it in-silico}                    &\bf Remarks\\\hline
%--------------------
Wild-Type	        & -			  & Low before attachment and high after & Low before attachment and high after& e.g. \cite{DeAntoni2005,Musacchio2007,Ibrahim2015Review,Ibrahim2008APC}\\\hline%
%%--------------------
Mad2		        & O 			& Low	        	                   & Low			                     & e.g. \cite{Kabeche2012}  \\
		            & D 			& High			                       & High		                       	 & e.g. \cite{Michel2001}   \\\hline
%%--------------------
BubR1:Bub3	        & O 			& Low		                           & Low 		  	                     & e.g.\cite{Yamamoto2007} \\
		            & D 			& High		                           & High			                     & e.g.\cite{Chan1999}     \\\hline
%%--------------------
Cdc20               & O 			& High                           	   & High			                     & e.g.\cite{Mondal2007}   \\
          		    & D 			&Low                            	   & Low			                     & e.g.\cite{Mondal2006}   \\\hline
%--------------------
\end{tabular}
\\[.2cm]
{D refers to \C{a} deletion or knockdown experiment, and O refers to an over-expression experiment.
In the simulations the respective initial concentration is set to
zero for the deletions, and \C{10-fold} higher concentrations for over-expression.}
}
%\label{tab:mutation}
\end{table}

\section*{acknowledgements}

This work was supported by the European \C{Commission} HIERATIC Grant 062098/14.
The author gratefully acknowledges the visiting fund of the Institute for Numerical Simulation (INS) at Bonn University.

\end{document}